# ANALYSIS OF KELVIN PROBE OPERATIONAL MODELS

**Eugeniu M. Popescu**[1]
Smithsonian Astrophysical Observatory
Harvard-Smithsonian Center for Astrophysics
60 Garden St., MS-63, Cambridge, MA 02138, USA

(May 23, 2011)

**Abstract**. We present a study of several models on which Kelvin Probe instruments with flat and spherical tips rely for operation and for the determination of the contact potential difference. Using covariance analysis, we have investigated the precision limits of each model as imposed by the Cramer-Rao bound. Where the situation demanded, we have evaluated the bias introduced by the method in the estimation of the contact potential difference.

[1] Email: epopescu@cfa.harvard.edu

## 1. INTRODUCTION

Many experiments such as the Laser Interferometric Space Antenna (LISA) [1], the Sounding Rocket Principle Of Equivalence Measurement (SR-POEM) [2] and the Satellite Test of the Equivalence Principle (STEP) [3] to name just a few, rely to a large extent on the use of capacitance gauges to control or monitor the position of test masses. The continuously increasing precision and accuracy requirements of these experiments impose very tight bounds on the magnitude of the residual electrostatic forces arising between test masses and capacitance gauge electrodes. These forces arise due to the inhomogeneity of the electrostatic potential across surfaces in close proximity and are commonly referred to as "patch effect" forces [4]. For illustration, the SR-POEM error budget requires that surface potential variations be below 0.1 mV over areas several $cm^2$ in size, and stable for durations of 40-100s [5].

Among the methods available for the investigation of the fabrication feasibility of such potential-uniform surfaces [6] is the Kelvin Probe (KP) method. In this method, the local contact potential difference (CPD) between a reference tip and a sample is determined from capacitance measurements of the oscillating tip in close proximity to the sample.

The performance of a KP instrument depends essentially on the particular operational model implemented in the software. As such, the design and construction of a performant instrument requires not only precise and accurate electronics, but also an operational model that ensures the highest performance of the instrument compatible with the hardware and with the experimental application.

The basic theoretical model underlying the operation of a KP instrument is quite well established at least for the case of flat tips [7]. Based on this theoretical model, several operational models have been developed and implemented on commercial and custom-built instruments for estimating the CPD and other parameters used for instrument control. The latter include the min-max probe-signal amplitude estimation model, the single and dual frequency Fourier amplitude estimation model and the amplitude estimation model by parabolic fitting.

In this paper we consider some of the most relevant KP operational models and we compare their performance as determined by standard covariance analysis methods. The paper is organized as follows. In Section 2, we review the elementary KP probe-signal models for flat and spherical tips, as well as the basics of the covariance analysis method. In Section 3, we present the covariance analysis of the full theoretical model, of the single and multiple frequencies Fourier projection model, and of the parabolic fitting model. We conclude the paper in Section 4 with a summary of the results and suggestions for future work.

## 2. THEORETICAL BACKGROUND

### 2.1 KP probe-signal models

KP instruments generally operate with two types of tips, namely flat and spherical. In the first case, the tip consists usually of a cylinder with diameter in the range 1mm – 1cm, with one of its flat circular faces forming a (circular) parallel-plate capacitor with the sample. In this case, the current through the tip is given by the expression:

$$i(t) = (V_B - V_{CPD})\frac{dC_F(t)}{dt} \tag{2.1.1}$$

with $V_B$ the backing potential, $V_{CPD}$ the contact potential difference between the tip and sample materials, and $C_F(t)$ the instantaneous tip-sample capacitance. Assuming that the tip and sample form an ideal parallel plate capacitor, the instantaneous capacitance is given by the standard expression $C_F = \varepsilon S/d(t)$ with $\varepsilon$ the dielectric permittivity of the medium between the tip and the sample, S the area of the circular face of the tip, and d(t) the instantaneous separation tip-sample separation. Furthermore, if the tip oscillates harmonically with respect to the sample, the instantaneous tip-sample separation can be expressed as $d(t) = d_0+d_1\cos(\omega t)$, with $d_0$ the (average) tip-sample separation, $d_1(< d_0)$ the oscillation amplitude and $\omega = 2\pi\upsilon$ the angular frequency of the oscillation. With these considerations, (2.1.1) yields for the flat probe current the expression:

$$i_P(t) = \frac{\omega\varepsilon S}{d_0}(V_B - V_{CPD})\frac{\gamma\sin(\omega t)}{[1+\gamma\cos(\omega t)]^2} \tag{2.1.2}$$

with $\gamma = d_1/d_0$ the modulation index. Equation (2.1.2) is the standard form of the flat tip KP operational model usually found in the literature [8], [9]. For future reference, the Fourier series decomposition of the probe current in (2.1.2) is given by the expression [8], [9]:

(2.1.3) $$\frac{\omega\varepsilon S}{d_0}(V_B\text{-}V_{CPD})\frac{\gamma\sin(\omega t)}{[1+\gamma\cos(\omega t)]^2}=\frac{\omega\varepsilon S}{d_0}(V_B\text{-}V_{CPD})\frac{2}{\sqrt{1\text{-}\gamma^2}}\sum_{k=1}^{\infty}(\text{-}1)^{k\text{-}1}k\left(\frac{\gamma}{1+\sqrt{1\text{-}\gamma^2}}\right)^k\sin(k\omega t)$$

For a spherical tip, a useful approximation of the tip-sample capacitance for d(t) << R is given by the expression [10]:

(2.1.4) $$C_S(t) = 2\pi\varepsilon R\left\{\ln\left[\frac{R}{d(t)}\right]+\ln(2)+\frac{23}{20}+\frac{\theta}{63}\right\}$$

where now d(t) is the separation between the apex of the tip and the sample, R is the tip radius and $\theta\in[0,1]$ time-independent constant. Inserting (2.1.4) into (2.1.1) yields for the spherical probe current the expression:

(2.1.5) $$i_S(t) = 2\pi\omega\varepsilon R(V_B\text{-}V_{CPD})\frac{\gamma\sin(\omega t)}{[1+\gamma\cos(\omega t)]}$$

The Fourier series decomposition of the probe current in (2.1.5) can be obtained through simple algebraic manipulations of the Fourier series for the flat probe current, and one has:

(2.1.6) $$2\pi\omega\varepsilon R(V_B\text{-}V_{CPD})\frac{\gamma\sin(\omega t)}{[1+\gamma\cos(\omega t)]} = 4\pi\omega\varepsilon R(V_B\text{-}V_{CPD})\sum_{k=1}^{\infty}(\text{-}1)^{k\text{-}1}\left(\frac{\gamma}{1+\sqrt{1\text{-}\gamma^2}}\right)^k\sin(k\omega t)$$

To the best knowledge of the author, this latter model has never been presented or discussed in the literature.

## 2.2 The covariance analysis method

In several experimental fields, e.g. in experimental gravity and experimental astrophysics, covariance analysis is a common tool for evaluating the performance limits of measurement systems in the presence of statistical noise. The domain of applicability of the method ranges from the analysis of simple systems like the ones in (2.1.2) and (2.1.5) to the analysis of extremely complicated physical systems like the SR-POEM, LISA and LIGO measurement systems [11]. For reasons of brevity, we will restrict the discussion of this method only to the models relevant to the scope of the paper, i.e. to the models in (2.1.2) and (2.1.5). For more detailed information, the interested reader is referred to [11] and the references within, as well as to the abundant literature available on the topic of optimal estimation, e.g. [12], [13].

Let $Y=\{Y_i \mid i = 1, \ldots, M\}$ be an experimental data set measured sequentially in time at equal time intervals $T=\{t_i \mid i=1,\ldots,M\}$. In our particular case, Y represents the output signal of the KP preamplifier acquired by the KP data acquisition electronics. Furthermore, assume that the measurement is corrupted by normally distributed zero-mean white noise with standard deviation $\sigma_0$. This assumption will be considered to hold identically over the entire dataset Y.

Also let $y(\alpha,\beta,\gamma|t)$ be a theoretical model depending parametrically on $\alpha,\beta,\gamma$ and piecewise smoothly on the time variable t. If this model is a candidate for the theoretical description of the data set Y, i.e. if the parameters $\alpha$, $\beta$, $\gamma$ are determined by a "best fit" procedure – $\chi^2$-fitting in this case – then the minimal statistical errors in the parameter estimation can be determined as follows. First, one constructs the Fisher information matrix F, which in our case is given by the expression:

$$(2.2.1) \qquad F = \frac{1}{\sigma_0^2}\begin{bmatrix} \sum_T \left(\frac{\partial y}{\partial \alpha}\right)^2 & \sum_T \left(\frac{\partial y}{\partial \alpha}\right)\left(\frac{\partial y}{\partial \beta}\right) & \sum_T \left(\frac{\partial y}{\partial \alpha}\right)\left(\frac{\partial y}{\partial \gamma}\right) \\ \sum_T \left(\frac{\partial y}{\partial \beta}\right)\left(\frac{\partial y}{\partial \alpha}\right) & \sum_T \left(\frac{\partial y}{\partial \beta}\right)^2 & \sum_T \left(\frac{\partial y}{\partial \beta}\right)\left(\frac{\partial y}{\partial \gamma}\right) \\ \sum_T \left(\frac{\partial y}{\partial \gamma}\right)\left(\frac{\partial y}{\partial \alpha}\right) & \sum_T \left(\frac{\partial y}{\partial \gamma}\right)\left(\frac{\partial y}{\partial \beta}\right) & \sum_T \left(\frac{\partial y}{\partial \gamma}\right)^2 \end{bmatrix}$$

The inverse $F^{-1}$ of the Fisher information matrix F contains the parameter variances along the principal diagonal and the parameter covariances (i.e. the unnormalized parameter correlation coefficients) in the off-diagonal positions. According to the Cramer-Rao theorem [11], the $F^{-1}$ is the lowest bound of the covariance matrices for the system – called the Cramer-Rao bound – giving hence the lowest possible variances and covariances of the parameter estimations.

Under these circumstances, the analysis of the KP operational models reduces to the calculation of the Cramer-Rao bound for each of the operational models under consideration and to the investigation of the behavior of the resulting variances and correlation coefficients for relevant values of the parameters $\alpha$, $\beta$, and $\gamma$.

## 3. ANALYSIS OF THE KP OPERATIONAL MODELS

Before proceeding any further, it is necessary to discuss a few very important details of our analysis. Generally, the software operation of a KP can be viewed as having two stages. In the first stage, one determines – according to the operational model implementation in the software – the amplitude of probe signal for two or more values of the backing potential $V_B$. Since according to the models presented in the previous section this amplitude is proportional to $V_B$, in the second stage the amplitude data as a function of the backing potential $V_B$ is fitted with a straight line. This fit yields $V_{CPD}$ as the abscissa intercept, as well as a second parameter (e.g. the slope of the fit line). This second parameter is generally a function of the tip-sample distance and is used to servo-control the latter.

A simple and straightforward covariance analysis of the second stage [14] allows one to develop the following strategy in choosing the values of $V_B$ for which the instrument should acquire data.

First of all, for a fixed amount of time per one $V_{CPD}$ measurement, it suffices for the instrument to acquire data in equal amounts of time for only two values of $V_B$. Commercial instruments have controls that allow the user to acquire data for one $V_{CPD}$ measurement for more than two values of $V_B$, but doing so for a fixed amount of time per $V_{CPD}$ measurement is equivalent to acquiring data for only two $V_B$ values for a longer time.

Secondly, the variance in the $V_{CPD}$ estimation is minimized if the two values of $V_B$ used in the data acquisition are symmetrical with respect to $V_{CPD}$, i.e. if one uses $V_{B1,2} = V_{CPD} \pm V_0$, with $V_0$ a suitably chosen voltage.

Under these circumstances, in all of the following it will be assumed that irrespective of the operational model implementation, a KP instrument operates by acquiring identical numbers M of data points equally spaced in time for only two values of the backing potential $V_B$ symmetrically placed with respect to $V_{CPD}$. Furthermore, as the KP probe signals are periodic for each value of $V_B$, it will also be assumed that data is acquired as described above during a single period for each value of the backing potential $V_B$.

Finally, there is also the issue of the phase delays associated with the start of the data acquisition for each value of the backing potential $V_B$. These phase delays can be included in the analysis as fitting parameters, but as it turns out, they have vanishing effects on the variances and covariances of the rest of the model parameters. Under these circumstances, for simplicity, we exclude these parameters from the present discussion and assume that the KP instrument is synchronized to start the data acquisition for each value of $V_B$ at zero phase delay.

**3.1 Analysis of the full KP operational models**

In all of the following, we will use the term "full" model to describe the operational models in which the probe signal acquired as described above is fitted with the theoretical models in (2.1.2) and (2.1.5) respectively. By contrast, all the other models that are approximations of the full models will be referred to as "partial" models.

With these considerations, for flat tips, the probe signal is fitted with a model of the form:

$$(3.1.1) \qquad y_P(t) = \alpha_P (V_B - \beta) \frac{\gamma \sin(\omega t)}{[1+\gamma\cos(\omega t)]^2}$$

in which the fitting parameters are defined as $\alpha_P = \omega\varepsilon GS/d_0$, $\beta = V_{CPD}$, and $\gamma = d_1/d_0$ with G the gain of the probe-current detection electronics. Similarly, for spherical tips, the probe signal is fitted with a model of the form:

$$(3.1.2) \qquad y_S(t) = \alpha_S (V_B - \beta) \frac{\gamma \sin(\omega t)}{[1+\gamma\cos(\omega t)]}$$

with $\alpha_S = 2\pi\omega\varepsilon GR$.

The Cramer-Rao bounds for the standard deviations of the model parameters and for their correlations are presented in Fig.1-Fig.4 below[2]. The correlations $\rho(\alpha\beta)$ and $\rho(\beta\gamma)$ are identically zero for the entire range of values of $\gamma$.

Several observations are in order at this point. First of all, it is only for flat tips that the full model allows complete control of the instrument based on the estimated values of the parameters. Indeed, one can use the estimated value of $\alpha_P$ (~ $1/d_0$) to control the tip-sample distance, and the ratio $\gamma/\alpha$ to control the tip oscillation amplitude. By contrast, for spherical tips, since $\alpha_S$ is independent of $d_0$, its estimated value cannot be used to feedback the tip-sample distance. However, this is not a major issue, since one can use the estimated value of $\gamma$ to control the tip-sample distance with the proviso that the tip oscillation amplitude be either generated by extremely precise and stable electronics, or servo-controlled separately and independently of the estimated values of the model parameters [15].

[2] All numerical results presented in this paper have been obtained using Wolfram Mathematica 8.0 (http://www.wolfram.com/).

Secondly – and this is a trend valid for all the operational model implementations discussed in this paper – the parameter standard deviations scale as $\sigma_0/ V_0$ for $\sigma(\alpha)$, $\sigma_0/\alpha$ for $\sigma(\beta)$ and $\sigma_0/\alpha V_0$ for $\sigma(\gamma)$ irrespective of tip shape. In particular this means that in order to reduce the Cramer-Rao bounds, one needs to operate the KP at high signal amplitudes $\alpha$ with reasonably large backing potentials $V_B$ relative to the measured $V_{CPD}$. This is, of course in accordance with the common practice in KP measurements.

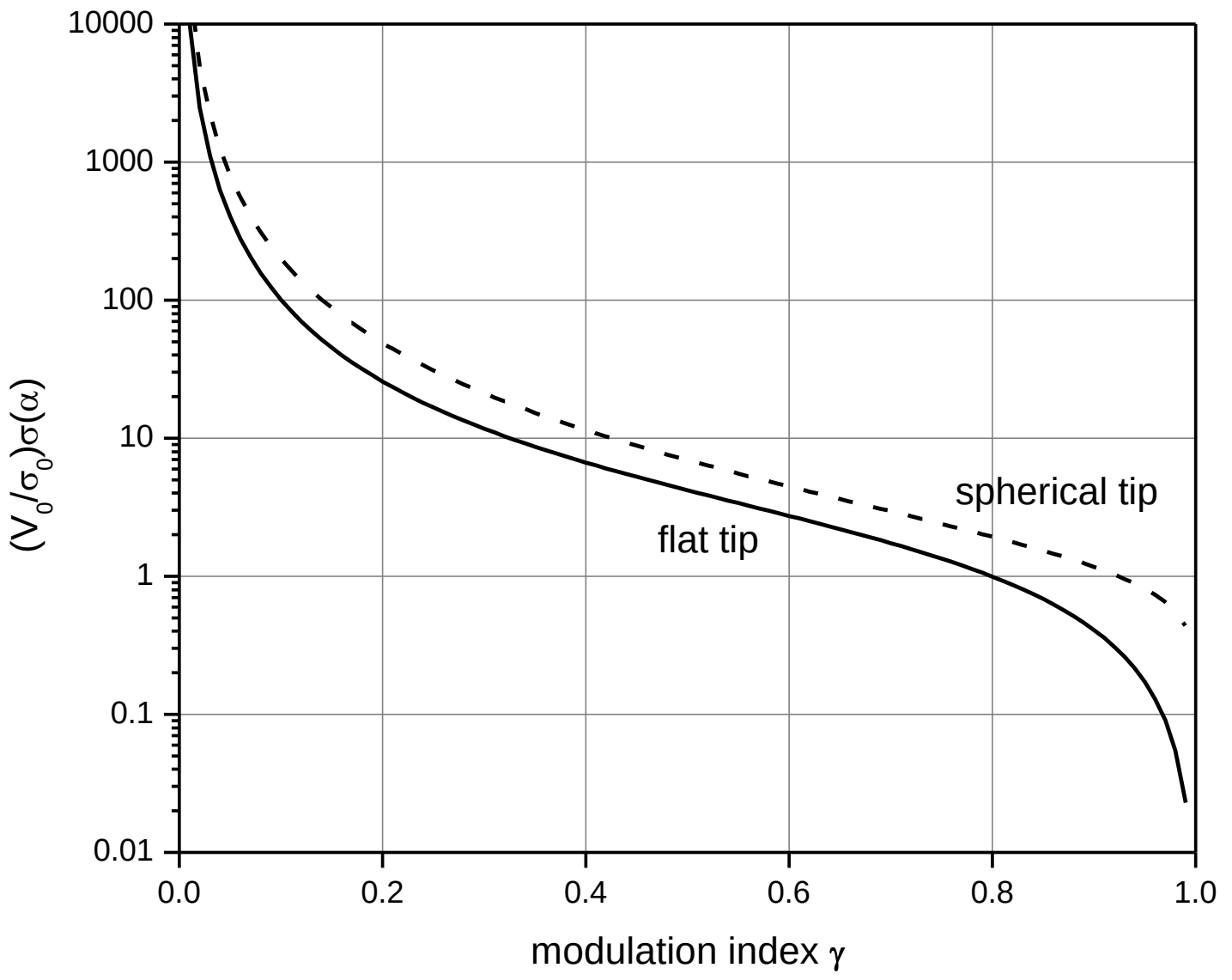

**Figure 1.** Scaled standard deviation for the full model estimation of the parameter $\alpha$.

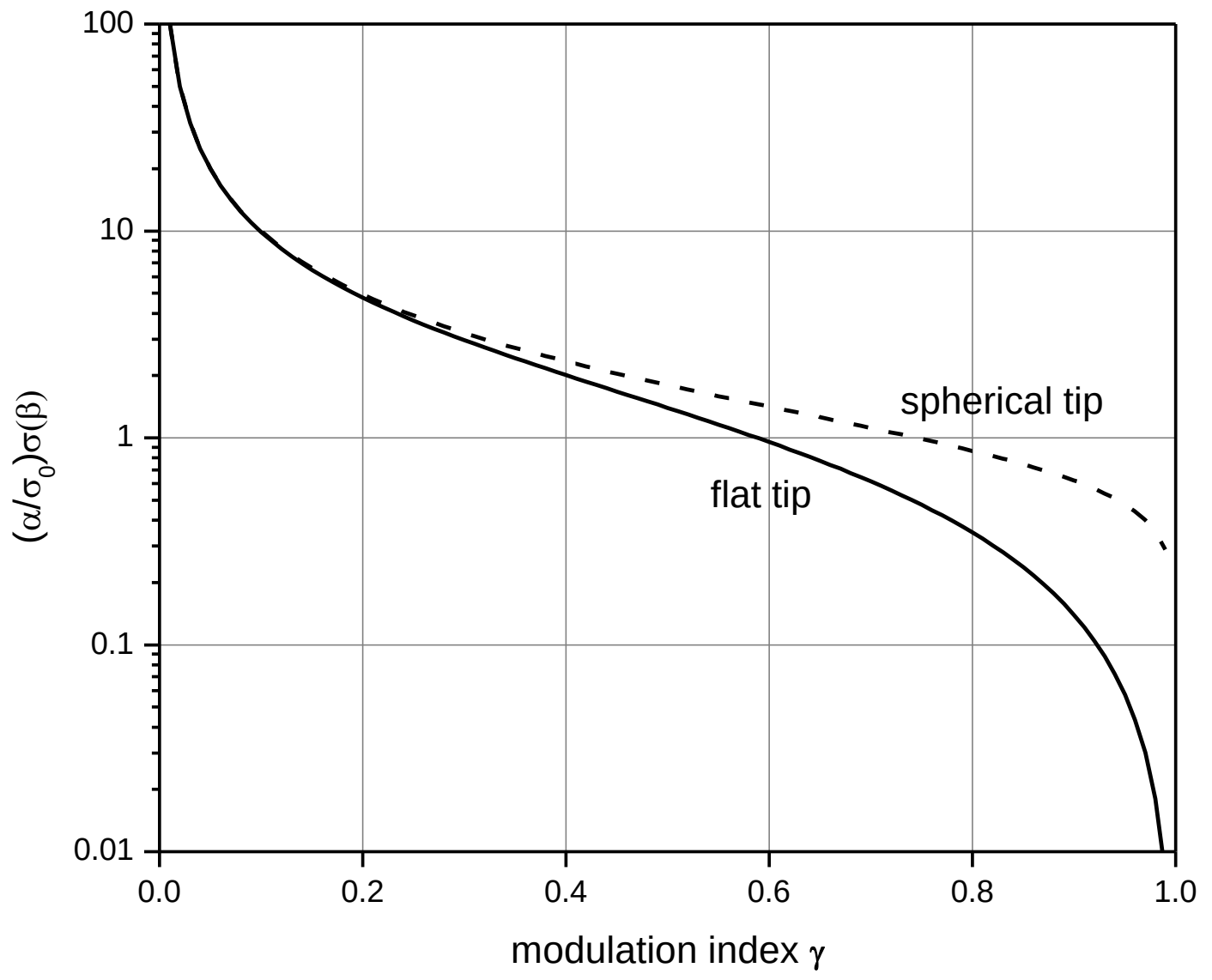

**Figure 2.** Scaled standard deviation for the full model estimation of the parameter β (≡$V_{CPD}$).

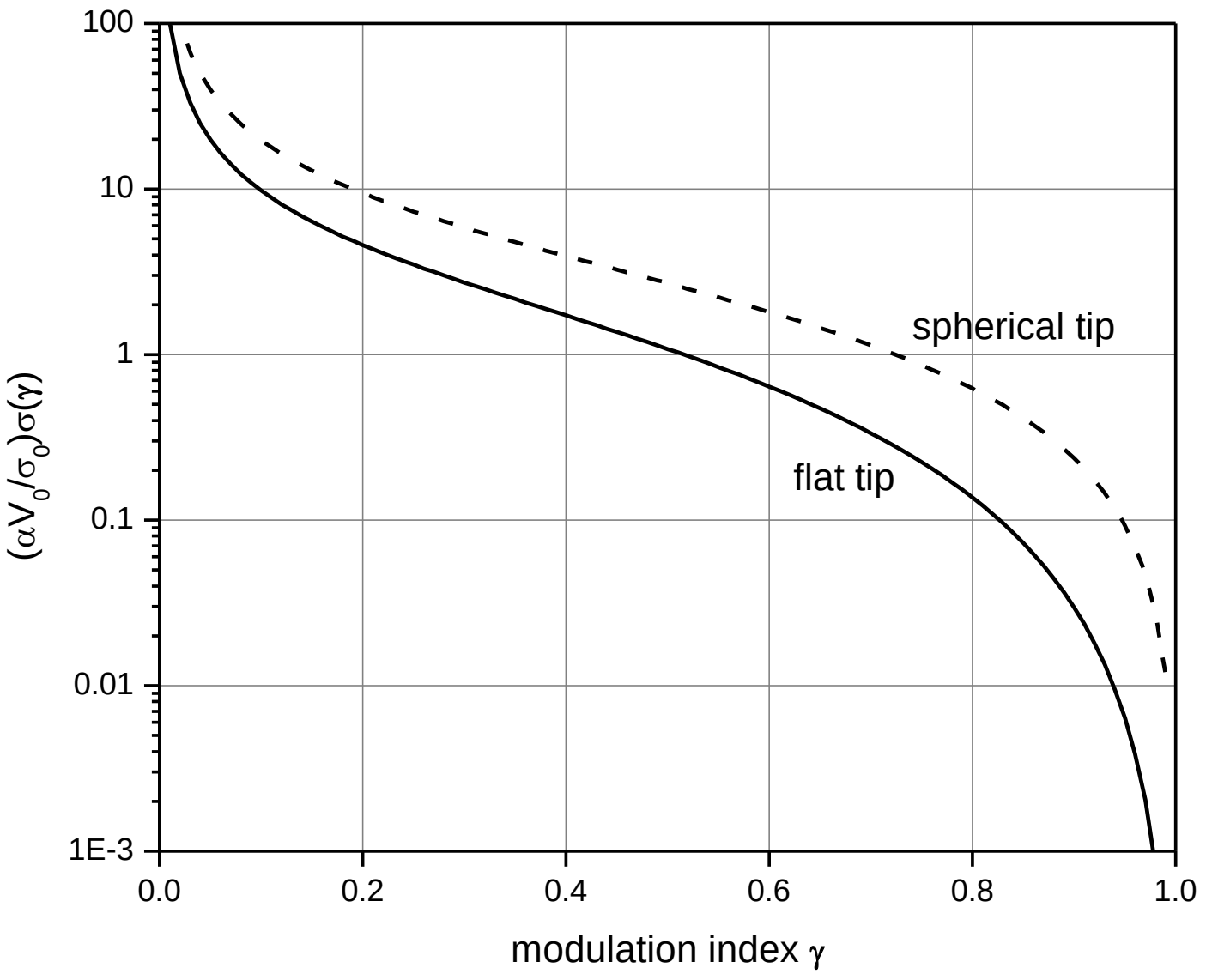

**Figure 3.** Scaled standard deviation for the full model estimation of the parameter γ.

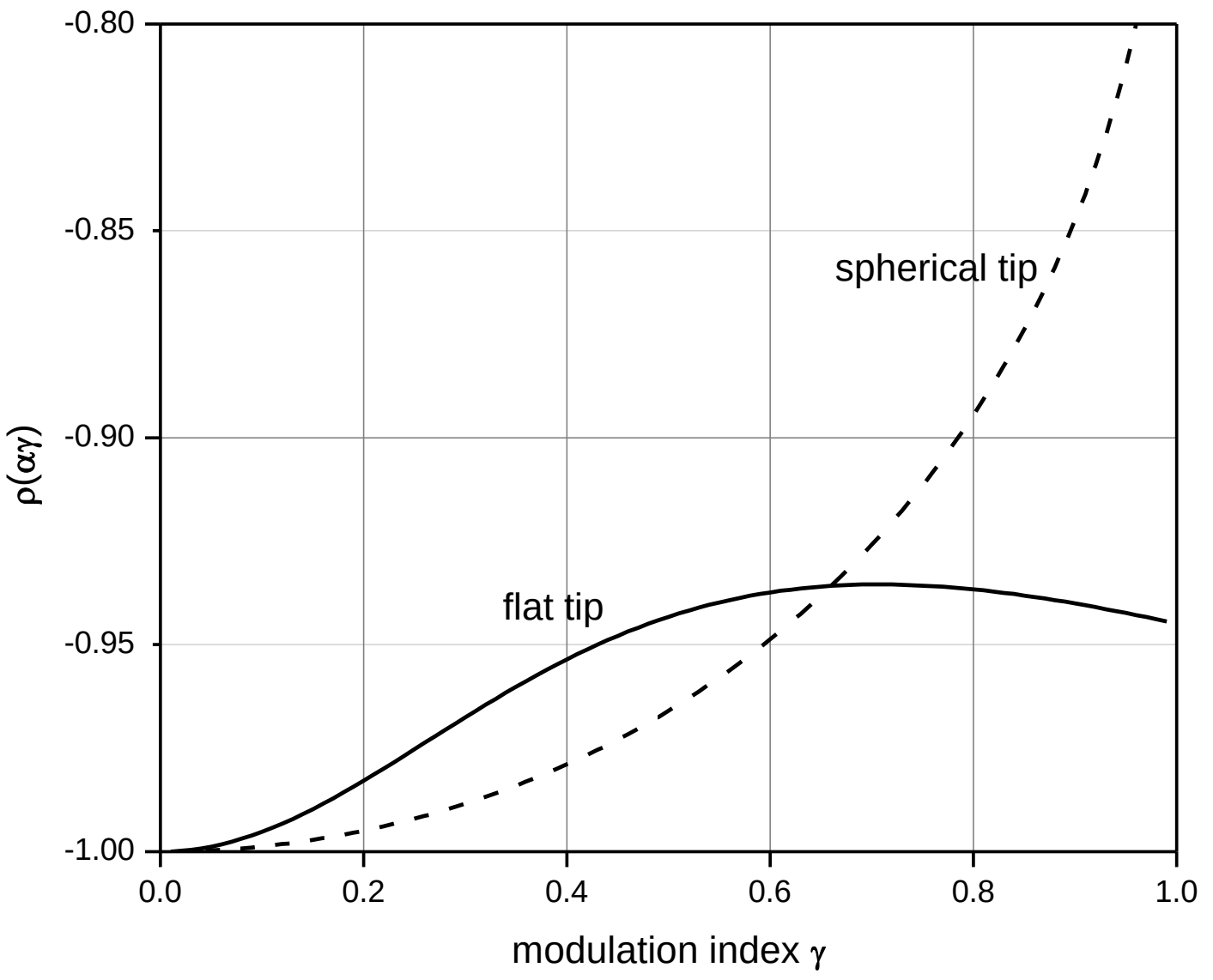

**Figure 4.** Correlation coefficient of the parameters α and γ in the full model covariance analysis.

Thirdly, it is clear from the expressions of $a_P$ and $a_S$ and from Fig.1-Fig.3 that for identical experimental conditions (same gain of the electronics, tip radius, tip-sample distance, oscillation amplitude and frequency) spherical tips yield poorer performance in terms of both signal amplitude and parameter standard deviations over the entire range of γ. However, they should not be discarded based solely on these conclusions, as they have several advantages over the flat tips. They are extremely easy to manufacture (e.g. by molten gravitational drop techniques) from a large variety of materials with extremely uniform surfaces and with radii of curvature in the range 0.5-5 mm. More importantly, they eliminate the need of precise alignment of the tip with the sample, which is a very difficult task especially for UHV KP instruments.

### 3.2 Analysis of the Fourier projection operational models

In the Fourier projection models, one uses as a convenient truncation of the Fourier series expansions in (2.1.3) and (2.1.6) , i.e. a projection of the probe-signal on a convenient subspace of the reciprocal Fourier space. As such, these models are clearly “partial” relative to the full models discussed in Section 3.1. Several Fourier projection operational models are in use [16], or have been or have been suggested in the literature [8]. To the best knowledge of the authors, all the models refer to flat tips only, and none of them use truncations of order larger than 2 of the corresponding Fourier series.

Consider the first order Fourier projection model. In this case, (2.1.3) and (2.1.6) yield for both types of tips fitting models of the form:

$$y_{P,S}(t) = A_{P,S}(V_B - \beta)\sin(\omega t) \qquad (3.2.1)$$

with $A_P = 2\alpha_P\gamma \Big/ \left(1-\gamma^2+\sqrt{1-\gamma^2}\right)$ and $A_S = 2\alpha_S\gamma \Big/ \left(1+\sqrt{1-\gamma^2}\right)$. However, for only two values of $V_B$ in the KP data acquisition settings, the models in (3.2.1) do not allow for the simultaneous estimation of all three parameters $\alpha$, $\beta$, $\gamma$, as the corresponding Fisher information matrix is degenerate, and hence cannot be inverted. Under these circumstances, the only possibility is to introduce the composite parameters $A_{P,S}$ and fit the data with the 2-parameter $(A,\beta)$ model in (3.2.1). The covariance analysis in this case yields for both types of tips $\sigma(A) = \sigma_0/ V_0$, $\sigma(\beta) = \sigma_0/ A$, and identically vanishing correlation between A and $\beta$. Similar to the results in the previous section, under identical experimental conditions, spherical tips yield poorer performance than flat tips. Control-wise, the situation is similar to the spherical tip case discussed in the previous section, where the tip-sample distance can be servo-controlled using the estimated value of the parameter A (~ to a function of $\gamma$) with the proviso that the oscillation amplitude is either generated with high precision and stability or is servo-controlled separately.

It should be noted that for $A_{P,S} = \alpha_{P,S}\gamma$, (3.2.1) also describes for the low modulation index approximation ($\gamma \ll 1$) of the full models in (2.1.2) and (2.1.5). As such, the above discussion remains equally valid for the latter approximations of the full models.

For $n^{th}$ order Fourier projection models with $n \geq 2$, the models contain the first n terms of the Fourier series and allow the numerical estimation of all three parameters $\alpha$, $\beta$, and $\gamma$ from the experimental data. The results of our covariance analysis for n = 2 and n = 10 are presented Fig.5-Fig.8 below. As previously, the correlations $\rho(\alpha\beta)$ and $\rho(\beta\gamma)$ vanish identically over the entire range of $\gamma$.

It is clear that Fig.5-Fig.7 show the same trend observed in the full model case in Section 3.1, namely that under identical experimental conditions, and irrespective of the order of the Fourier projection model, the performance of spherical tips is lower than that of flat tips.

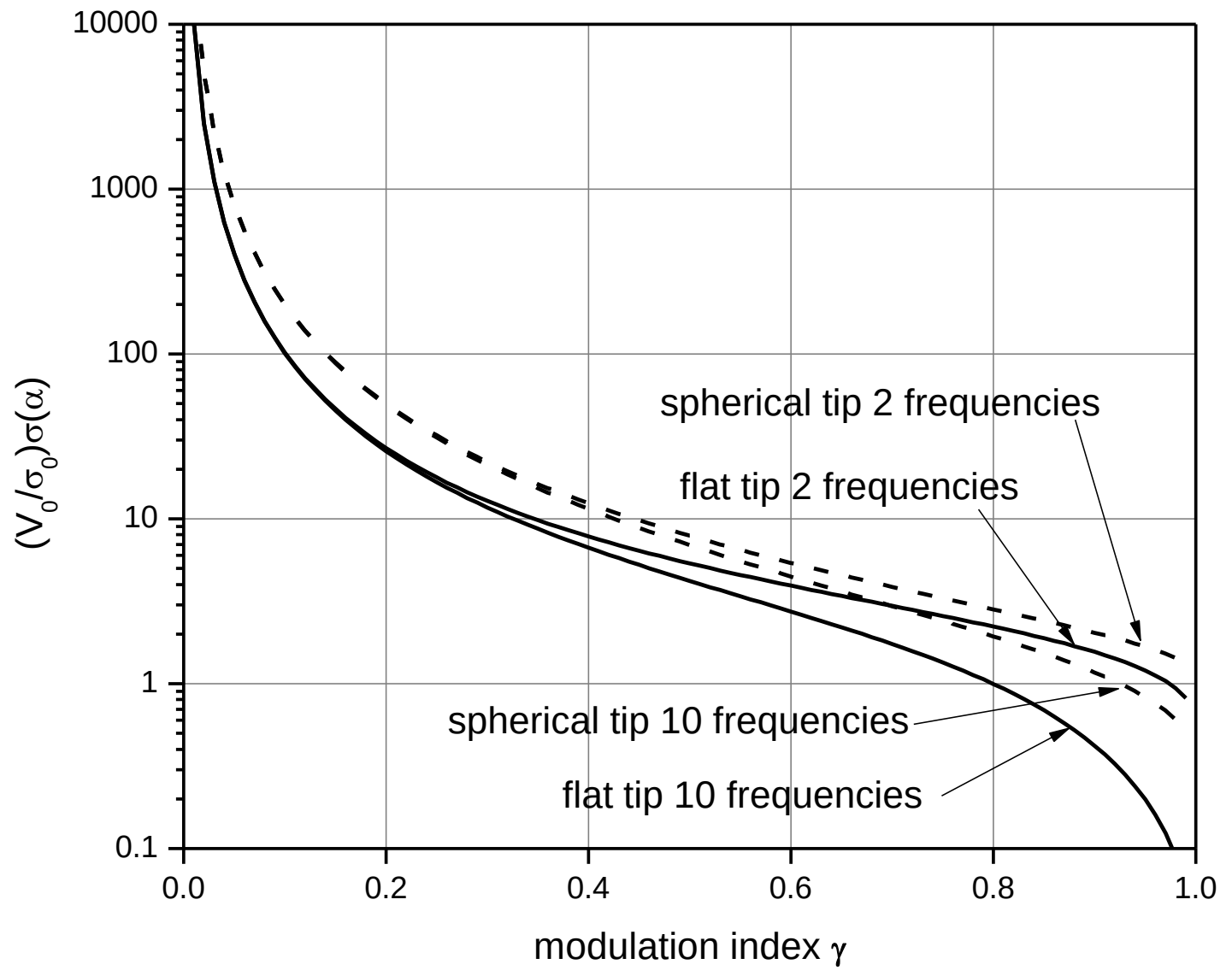

**Figure 5**. Scaled standard deviation for the Fourier projection estimation of the parameter $\alpha$.

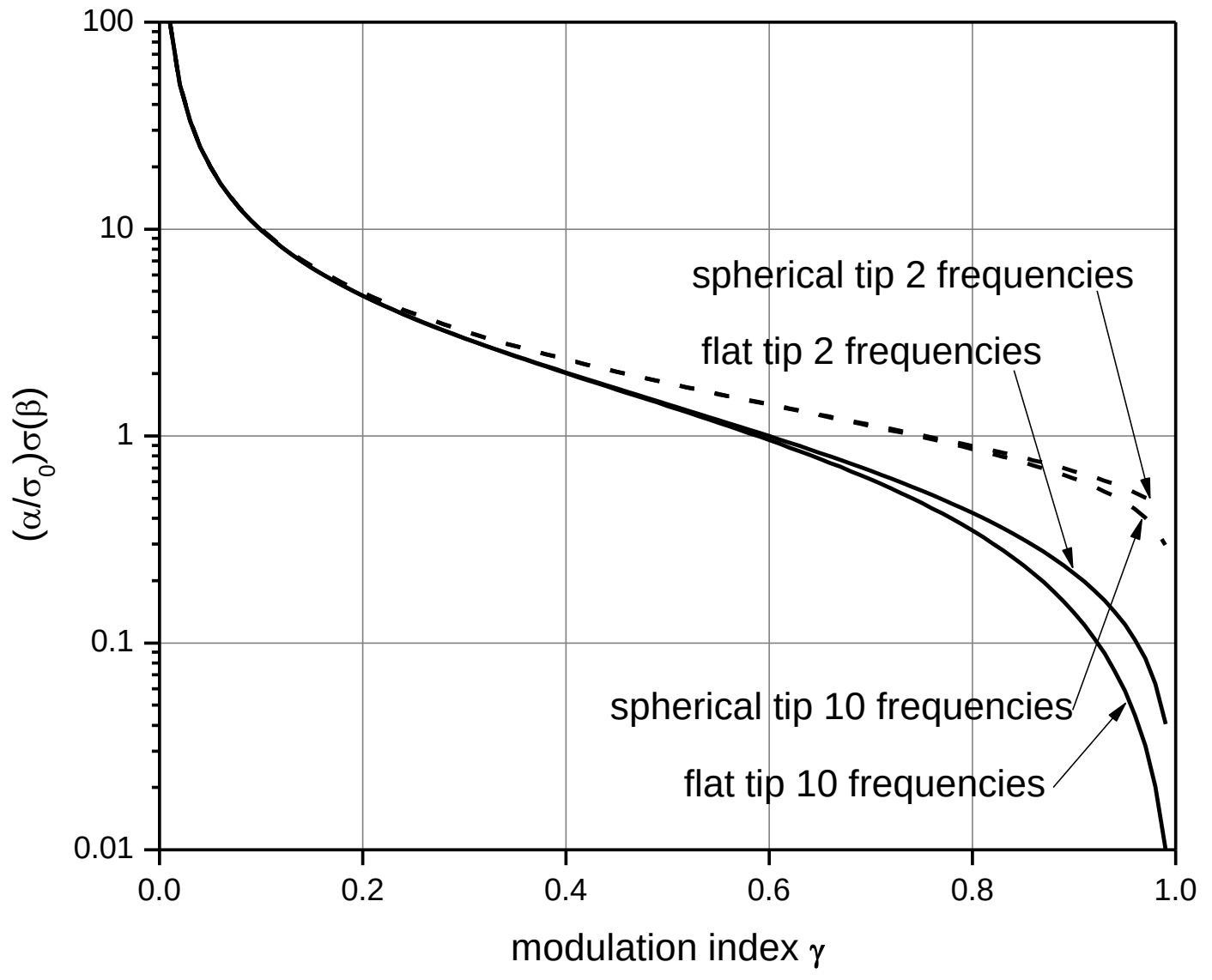

**Figure 6**. Scaled standard deviation for the Fourier projection estimation of the parameter $\beta(\equiv V_{CPD})$

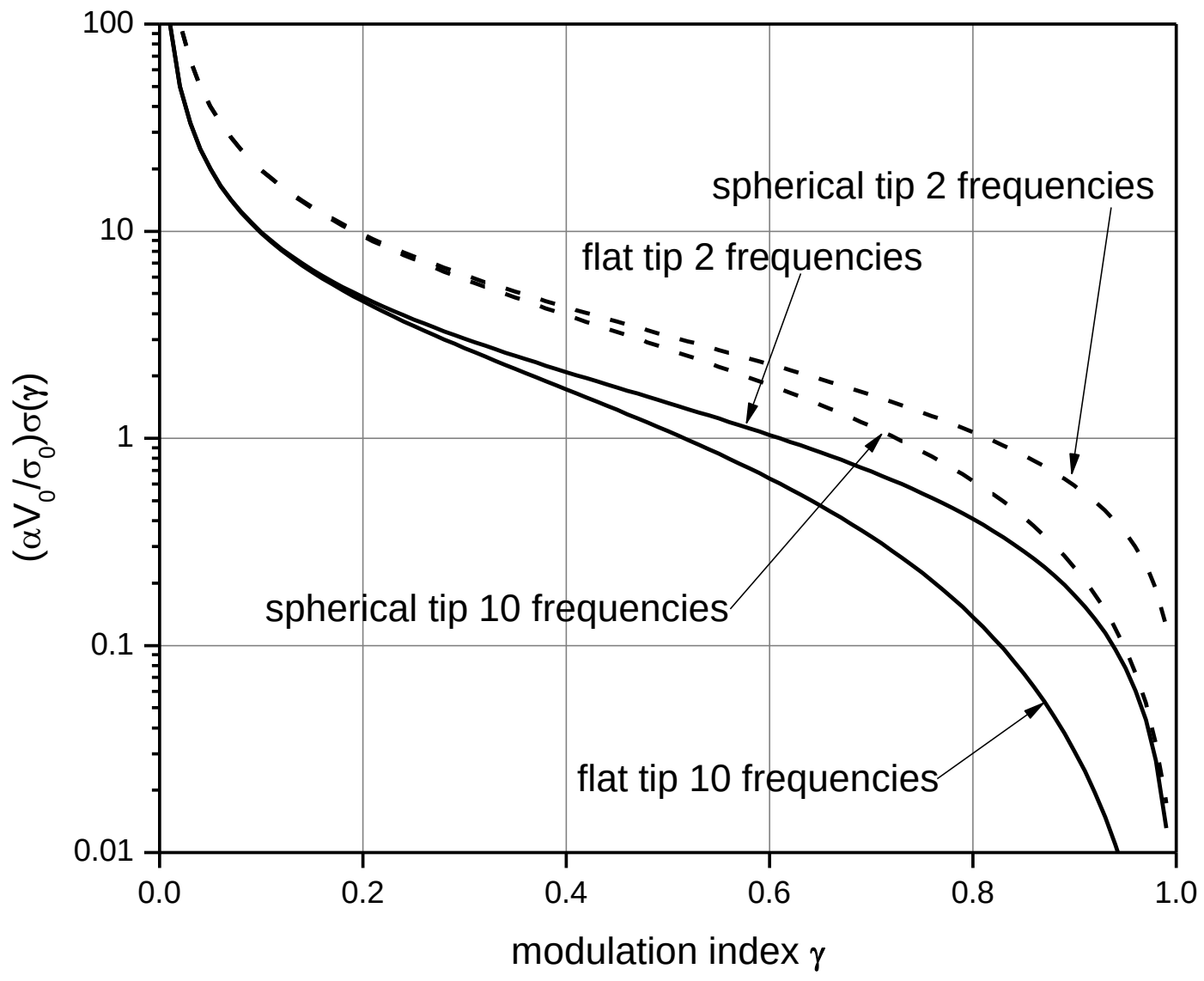

**Figure 7**. Scaled standard deviation for the Fourier projection estimation of the parameter $\gamma$.

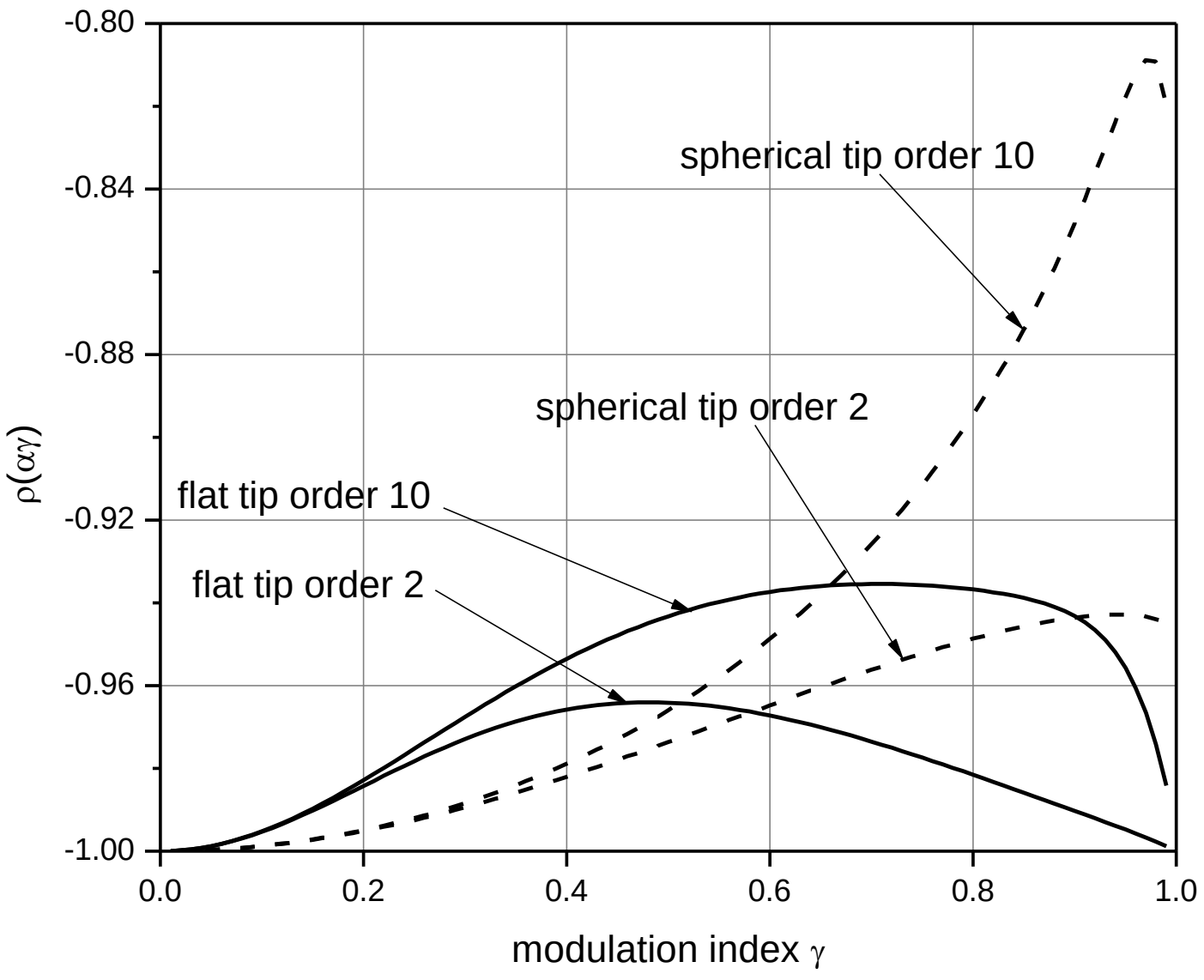

**Figure 8**. Correlation coefficient of the parameters $\alpha$ and $\gamma$ in the Fourier projection model covariance analysis.

And similar to the full model case, only the flat tip models allow for the full control of the instrument using the estimated values of the parameters $\alpha$, $\beta$, and $\gamma$, while the spherical tip models require either precise and stable generation or independent servo-control of the tip oscillation amplitude in order to achieve full control of the instrument. Also similar to full model case, the estimated value of the parameter β (≡ $V_{CPD}$) is uncorrelated to the values of α and γ, while the latter two are strongly anti-correlated as shown by Fig.8.

It should be noted that by increasing the order of the Fourier projection model, the performance of the model increases in both tips configurations. However, in practice the maximum order of the models is limited to n=10÷20 by the noise floor of the instrument, and going beyond this limit only adds to computational complexity with minimal improvements in model performance.

Finally, as the Fourier projection models are partial models, this raises the issue of the bias introduced by these models (relative to the full model in Section 3.1) in the estimation of the parameters α, β, and γ. Our study shows, surprisingly perhaps, that the estimation biases for all parameters and for both types of tips are negligibly low, below $10^{-13}$ % for n = 2 and decreasing slightly with increasing n.

**3.3 Analysis of quadratic fitting operational models**

The KP probe signal for both flat and spherical tips is periodic in time, and over each period it exhibits two extrema whose shapes depend on the values of γ as illustrated in Fig.9. Quadratic fitting models are partial models that use second-order time-series expansions of the full models in (3.1.1) and (3.1.2) around their maxima and minima to estimate the parameters α, β, and γ from the peak-to-peak amplitudes of the probe-signal for the two values of the backing potential $V_B$.

Consider first the flat tip model in (3.1.1). Taking the time reference at t = 0, the maxima and minima of the probe signal occur at the time coordinates:

(3.3.1) $$t_{P|max(-)/min(+)}(k) = \frac{1}{\omega}\left[\left(2k + 1\right)\pi \mp \arccos\left(\eta\right)\right]$$

with k integer and with η defined as:

(3.3.2) $$\eta = -\frac{1}{2\gamma} + \sqrt{2 + \frac{1}{4\gamma^2}}$$

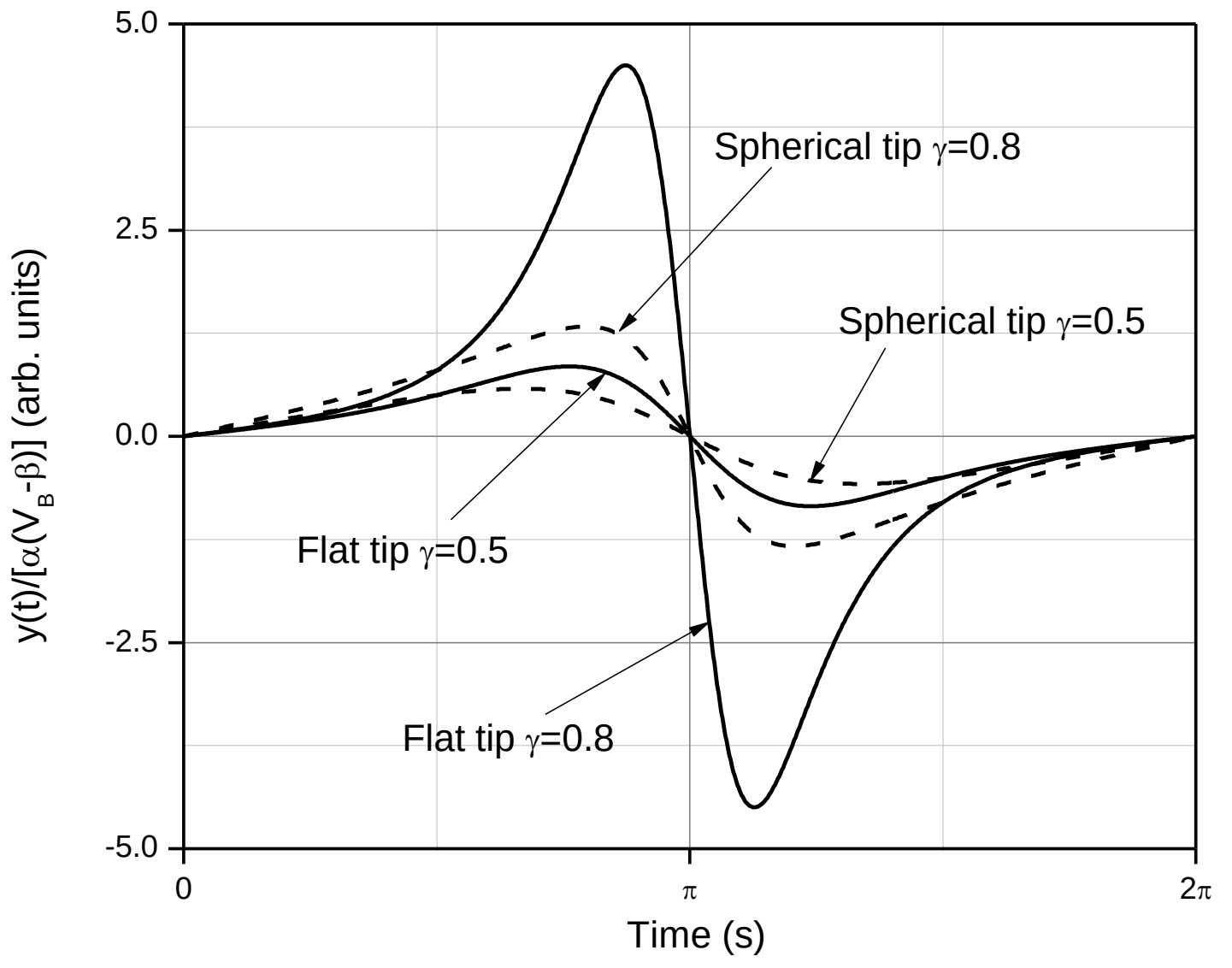

**Figure 9**. Scaled probe signal for $\gamma$ = 0.5 and $\gamma$ = 0.8.

Under these circumstances, the second-order expansion of (3.1.1) around $t_{max}$ and $t_{min}$ will have the expression:

(3.3.3) $$y_P\left[t|t_{P|max(-)/min(+)}(k)\right] = \mp\frac{\alpha_P\gamma\sqrt{1-\eta^2}}{(1-\gamma\eta)^2}(V_B-\beta)\left\{\frac{\omega^2\delta}{2(1-\gamma\eta)^2}\left[t-t_{P|max(-)/min(+)}(k)\right]^2-1\right\}$$

with $\delta$ defined as:

(3.3.4) $$\delta=1+4\beta\gamma+\beta^2\gamma^2-6\gamma^2$$

Similarly, for the spherical tip model in (3.1.2), the maxima and the minima of the probe signal occur at the time coordinates:

(3.3.5) $$t_{S|max(-)/min(+)}(k)=\frac{1}{\omega}\left[(2k+1)\pi\mp\arccos(\gamma)\right]$$

and the second order expansion of (3.1.2) around these points will be given by the expression:

(3.3.6)
$$y_S\left[t|t_{S|max(-)/min(+)}(k)\right] = \mp\frac{\alpha_S\gamma}{\sqrt{1-\gamma^2}}\left(V_B - \beta\right)\left\{\frac{\omega^2}{2\left(1-\gamma^2\right)}\left[t - t_{S|max(-)/min(+)}(k)\right]^2 - 1\right\}$$

The performance of the quadratic fitting method depends on the goodness of the fit of the probe-signal extrema. As such, it depends essentially on the choice of datapoints around the extrema that are used for the fit. To quantify the effects of the latter, we have kept the number of datapoints per probe-signal period constant, and have introduced an empirical parameter $f\in[0,1]$ that serves as a measure of the number of datapoints in the interval $[t_1(fy_{max}), t_2(fy_{max})]$ centered around $t_{P,S|max(-)}(k)$. This number of points is used for the fitting of both the maximum and the minimum of the probe-signal in the same oscillation period.

The results of our analysis are presented in Fig.10-Fig.16 for $f = 0.2, 0.4, 0.6, 0.8$. Only the $f = 0.2$ and $f = 0.8$ curves are marked explicitly, while the rest are displayed between the marked curves in sequential ascending (Fig.10-Fig.13) and respectively descending (Fig.14-Fig.16) order.

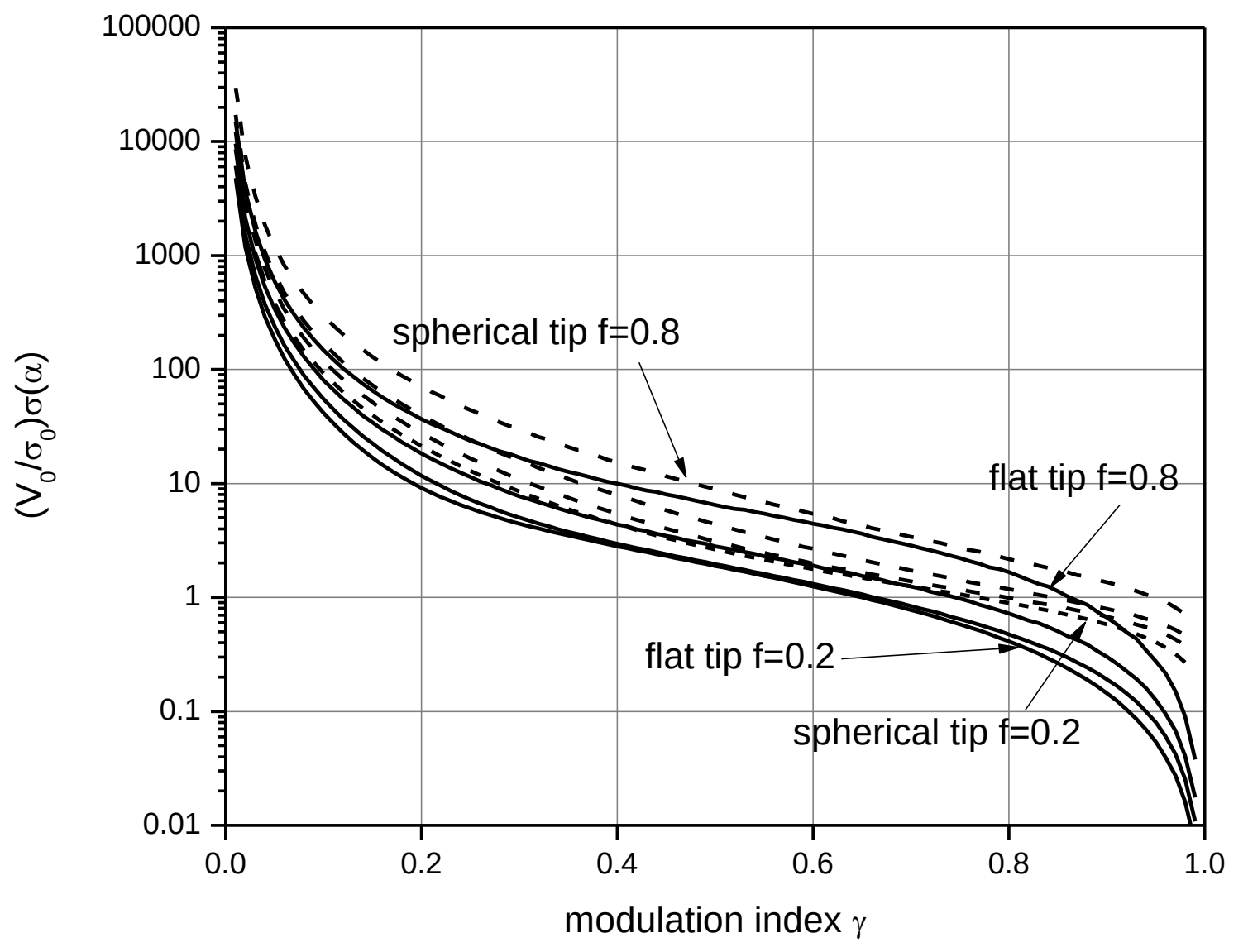

**Figure 10**. Scaled standard deviation for the quadratic fit estimation of the parameter $\alpha$.

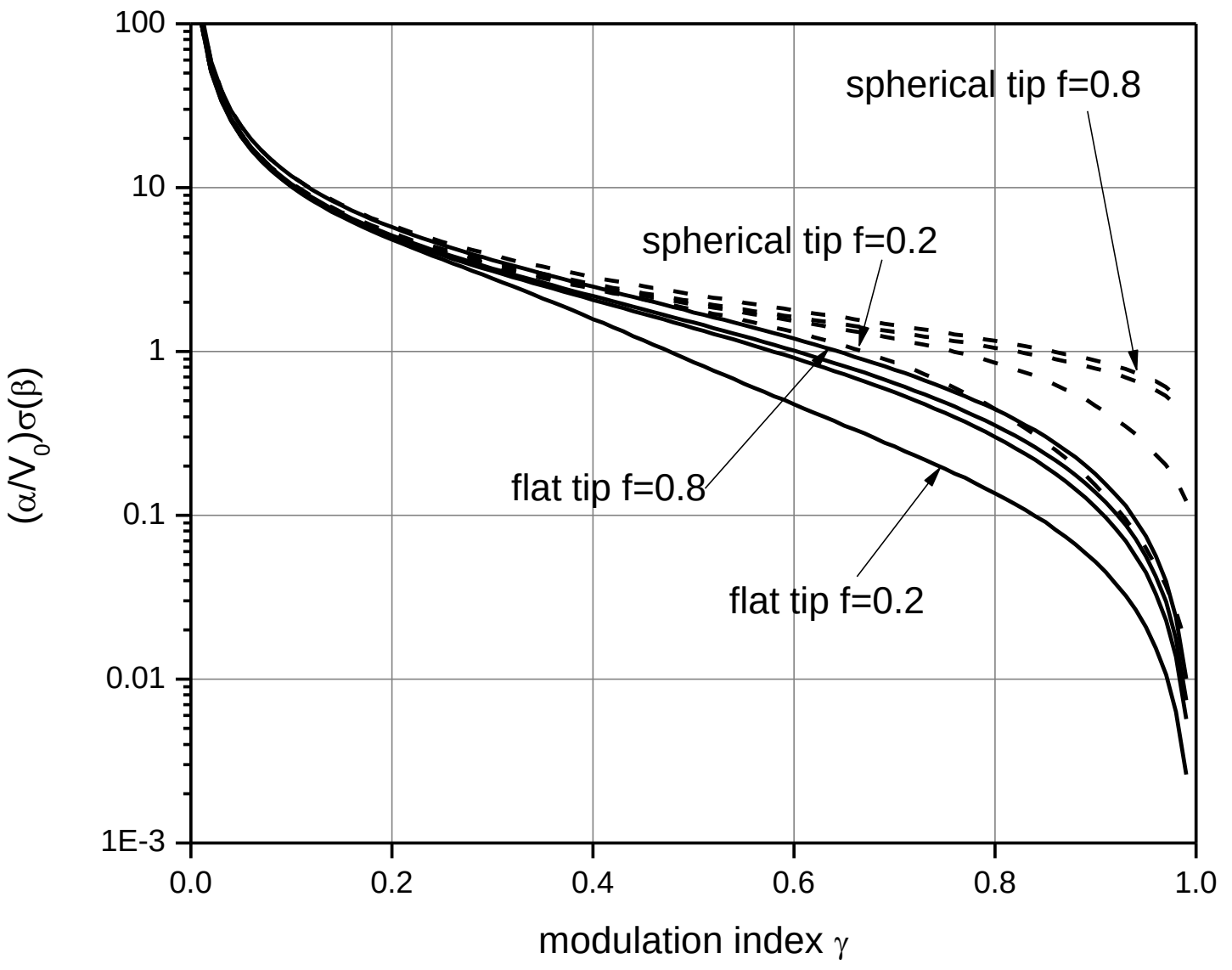

**Figure 11**. Scaled standard deviation for the quadratic fit estimation of the parameter $\beta(\equiv V_{CPD})$.

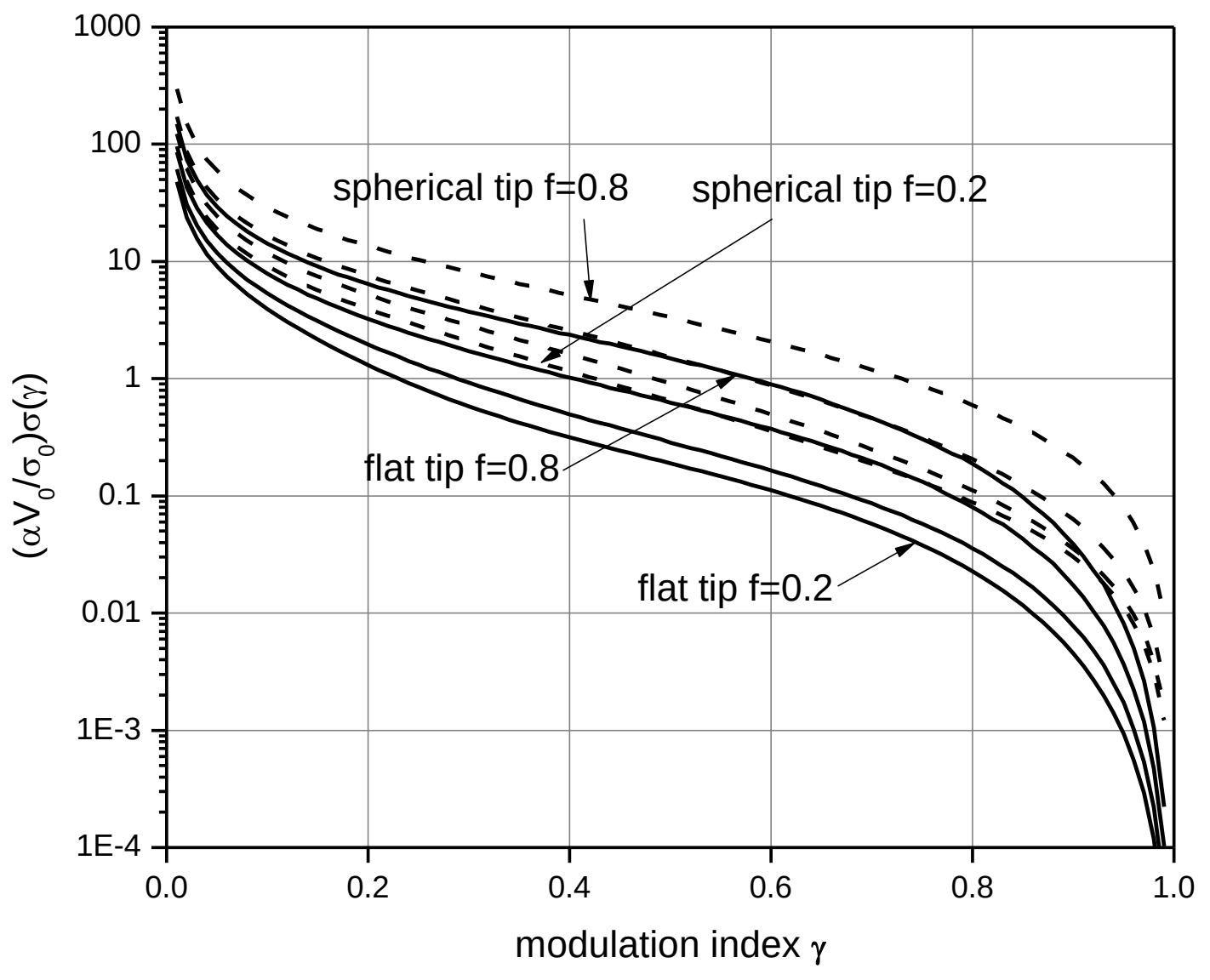

**Figure 12**. Scaled standard deviation for the quadratic fit estimation of the parameter $\gamma$.

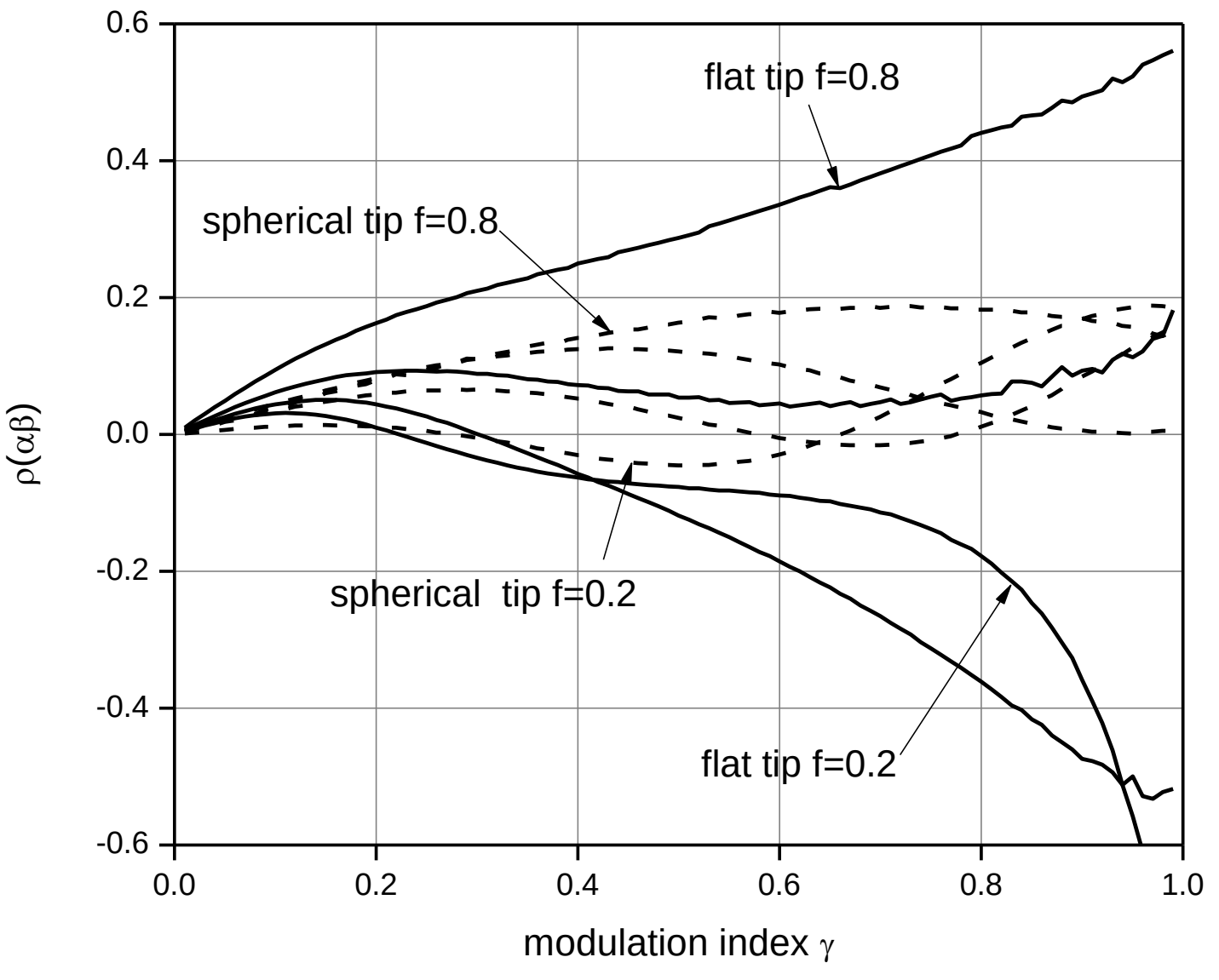

**Figure 13**. Correlation coefficient of the parameters α and β in the quadratic fit model covariance analysis.

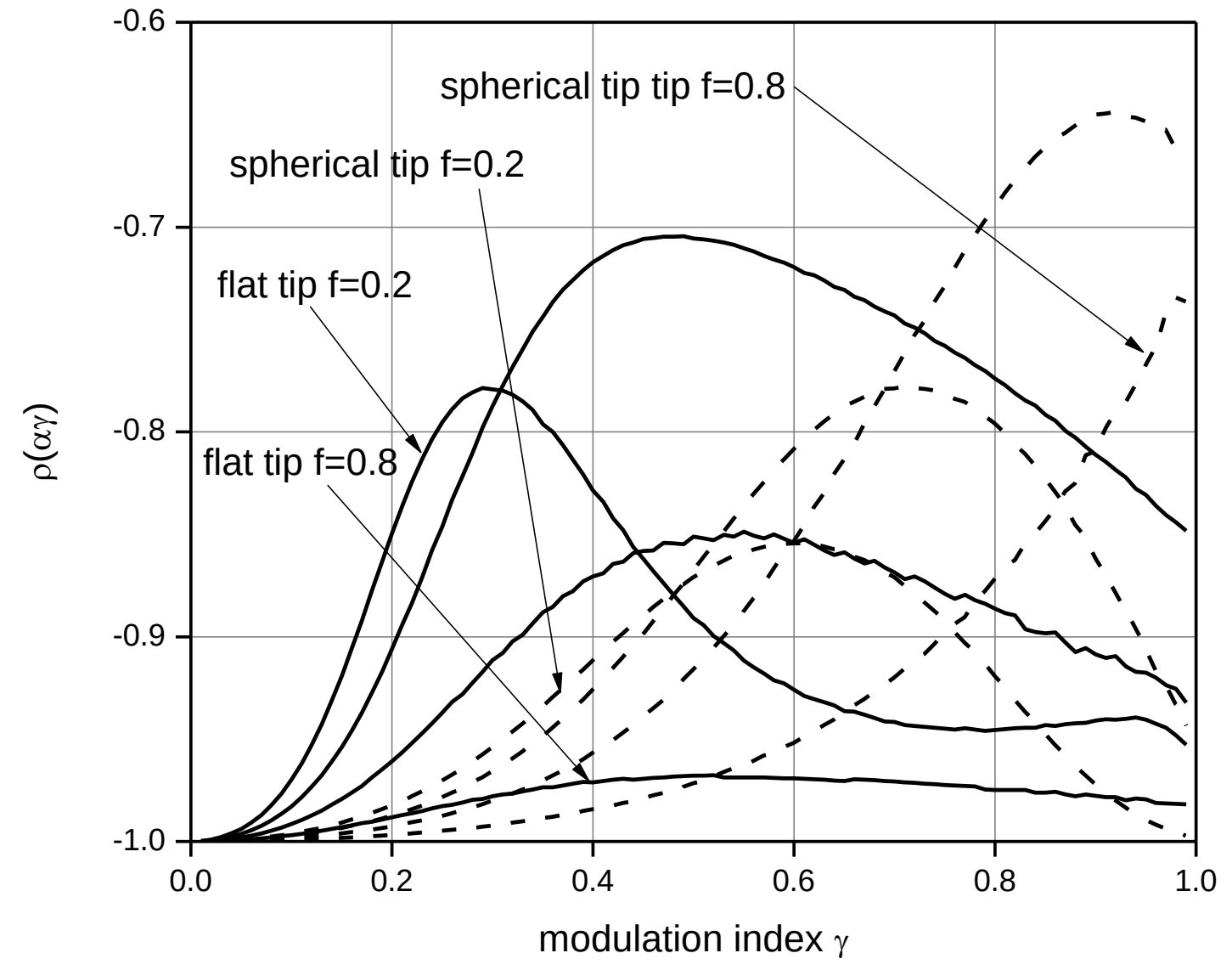

**Figure 14**. Correlation coefficient of the parameters α and γ in the quadratic fit model covariance analysis.

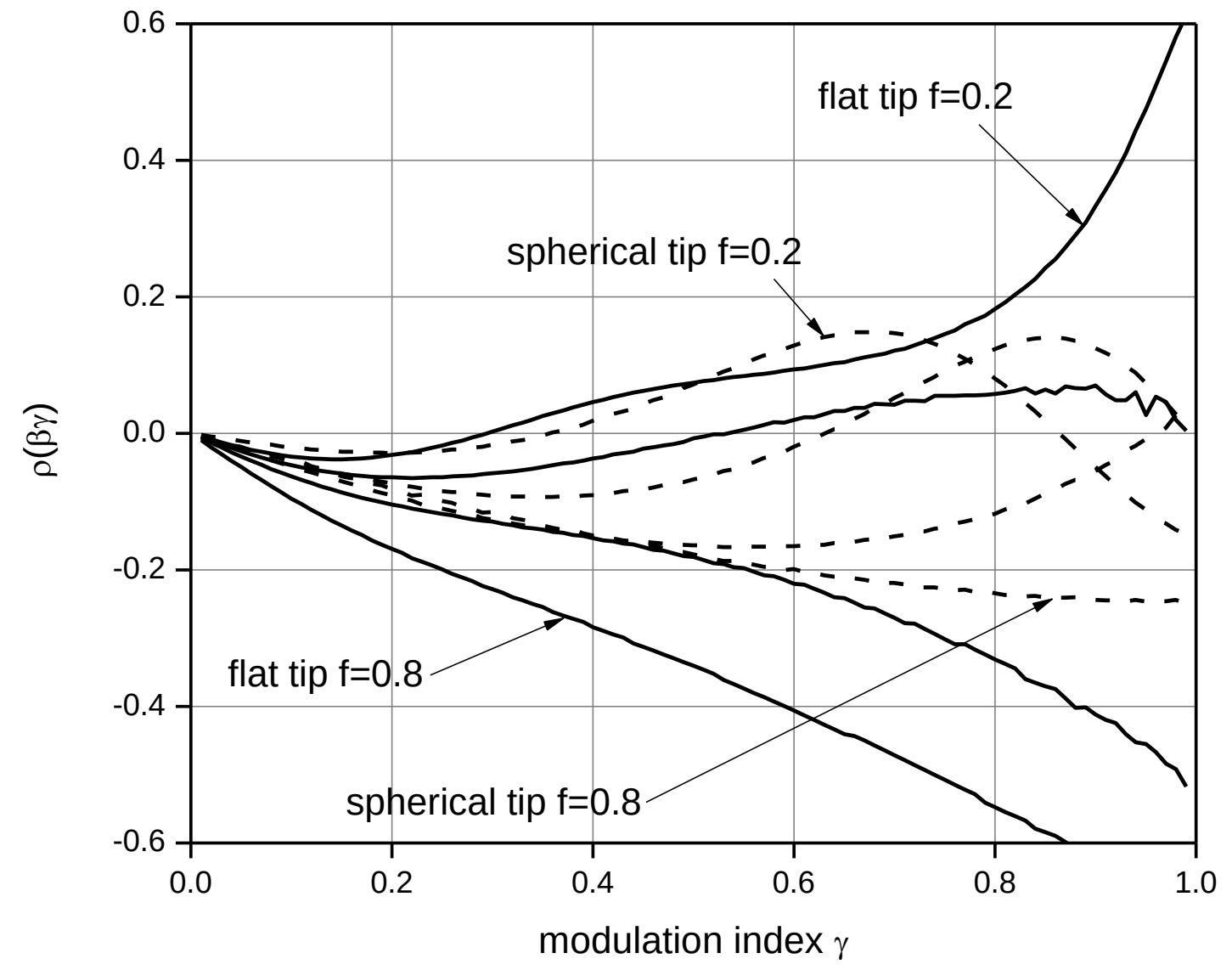

**Figure 15**. Correlation coefficient of the parameters β and γ in the quadratic fit model covariance analysis.

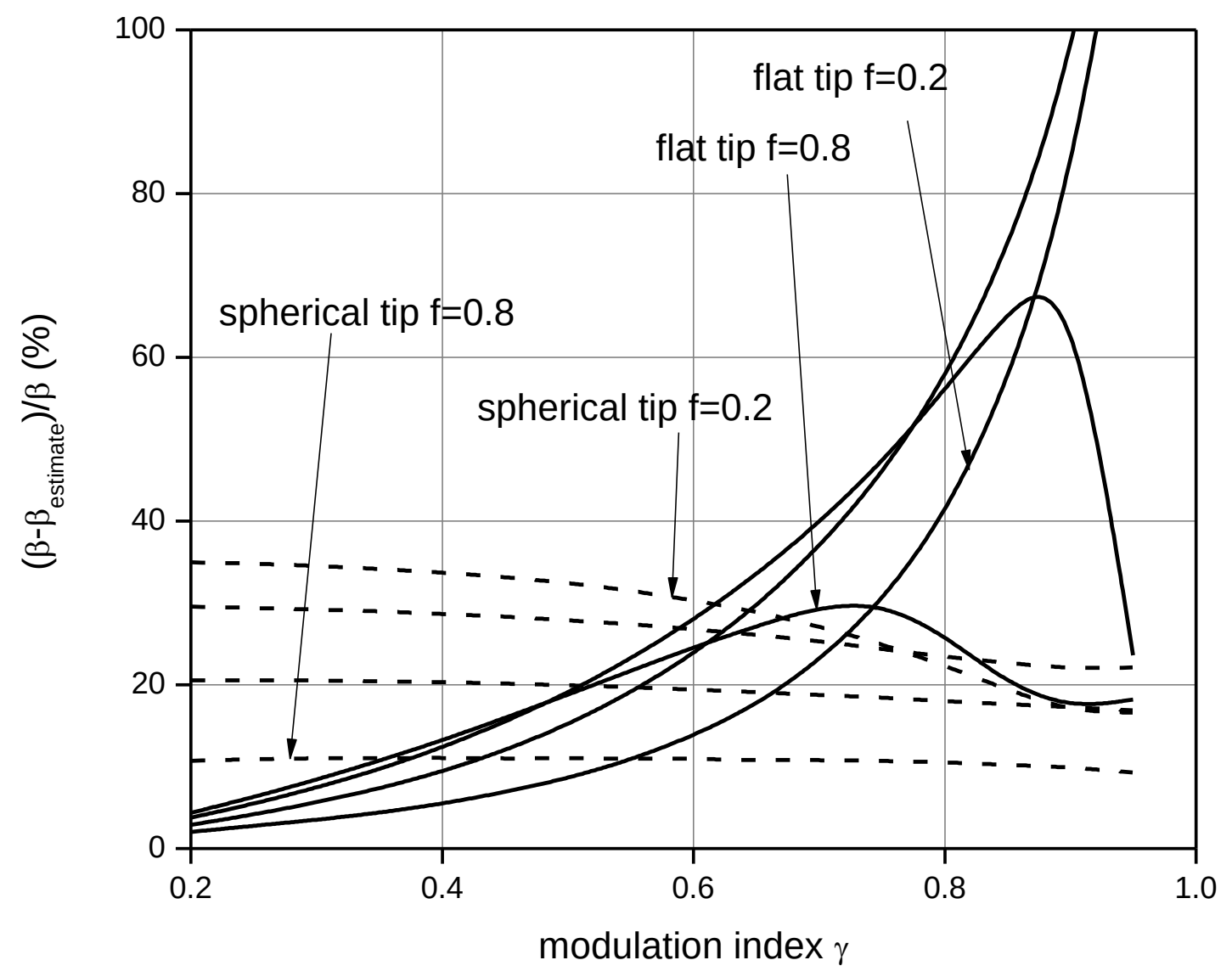

**Figure 16**. The bias introduced in the estimated values of the parameter $\beta(\equiv V_{CPD})$ by the quadratic fitting model relative to the full model.

Much like in the previous cases, it is clear from Fig.10-Fig.12 that under identical experimental conditions spherical tips exhibit poorer performance than flat tips. These graphs also show that the performance of the model increases with increasing modulation index $\gamma$ and decreasing value of the parameter f. However, in contrast to the other models, the quadratic fitting model exhibits non-vanishing correlations between the parameter $\beta$ ($\equiv V_{CPD}$) and the parameters $\alpha$ and $\gamma$, as shown by Fig.13 and Fig.15.

Also in contrast to the other models, the quadratic fitting models introduce very large biases in the estimation of the parameters. For illustration, in Fig.16 we present a study of the bias in the parameter $\beta$ ($\equiv V_{CPD}$) introduced by the quadratic fitting model relative to the full model in Section 3.1 with parameters $\alpha = 1$, $\beta = 0.35$.

As it can be seen from this figure, for flat tips, the bias in estimating $\beta$ ($\equiv V_{CPD}$) is lowest for low values of $\gamma$. For low values of f, the bias increases with increasing $\gamma$ to more than 100%, but a marked decrease in bias occurs for large values of these parameters. For spherical tips, the biases are larger for smaller values of $\gamma$ and decrease with increasing f and $\gamma$. For large values of these parameters, spherical tips introduce smaller biases than flat tips.

## 4. SUMMARY AND CONCLUSIONS

To summarize, we have investigated three major KP operational models, namely the full analytic model, the n-order Fourier projection model with $n \geq 1$, and the quadratic fitting model for both flat and spherical tips.

Our analysis shows that flat tip models – with the exception of the $1^{st}$-order Fourier projection model – allow for the full control of the instrument based on the estimated values of the model parameters. By contrast, spherical tip models require for the same task either highly precise and stable hardware for the generation of the tip oscillation amplitude, or additional detection hardware for the separate servo-control of the latter. Furthermore, our analysis shows that under identical experimental conditions (same probe-signal gain, same tip-sample spacing, same oscillation amplitude and same values of the backing potential $V_B$), the performance of spherical tips is systematically lower than that of flat tips.

In terms of model performance, our analysis shows that full models are the most performant, with low estimation errors of the model parameters and vanishing correlations between $V_{CPD}$ and the rest of the model parameters. The n-order Fourier projection models with $n \geq 2$ follow closely, with negligible bias but larger errors in the parameter estimation. The quadratic fitting model yields low parameter estimation errors, but unfortunately exhibits significant biases in the estimated values of all model parameters.

From the viewpoint of the requirements imposed on the performance of such an instrument by the SR-POEM experiment, the present study has clarified in detail the options available for the static operation and control of a KP instrument. However, as mentioned in Section 1, we need an instrument capable of scanning surfaces several $cm^2$ in size with mm-order resolution in times of the order of 20-50s. Unfortunately, currently available KP instruments are unable to perform such fast scans with enough accuracy and reliability. Among the reasons behind this inability is the lack of fast and precise instrument control, which prevents the instrument from scanning smoothly over the sample without disengaging and re-engaging servo-control. This problem constitutes the topic for our next study, and we will report on it in the near future.

**ACKNOWLEDGEMENTS**

This work was supported by the NASA grant NNX08AO04G. The author would like to thank Robert D. Reasenberg and Bijunath R. Patla (SAO-CfA) for useful discussions and suggestions, as well as James P. Cowin (PNNL) for in-depth discussions on the operation of KP instruments.

**REFERENCES**

1. Danzmann, K., Class. Quant. Grav. **13**, A247-A250 (1996).
2. Reasenberg, R. D. et al, Class. Quant. Grav. **28**, 094014 (2011).
3. Sumner, T. J. et al, Adv. Space Res. **39**(2), 254-258 (2007).
4. Speake, C. C., Class. Quant. Grav. **13**, A291-A297 (1996).
5. Patla, B. et all, BAPS.2010.APR.S10.6.
6. Robertson, N. A. et al., Class. Quant. Grav. **23**, 2665-2680 (2006).
7. Baikie, I. D., "Old Principles…New Techniques: A Novel UHV Kelvin Probe And Its Application In The Study Of Semiconductor Surfaces", Enchede, The Netherlands, 1988.
8. Baumgartner, H., Meas. Sci. Technol. 3, 237-238 (1992).
9. Mackel, R., Baumgartner, H., Ren, J., Rev. Sci. Instr. **64**(3), 694-699 (1992).
10. Boyer, L. et al, J. Phys. D: Appl. Phys. **27**, 1504-1508(1994).
11. Vallisneri, M., Phys. Rev. D **77(4)**, 042001 (2008) .

12. Gelb, A. (editor), “Applied Optimal Estimation”, MIT Press (1974).
13. Edwards, Robert V., “Processing Random Data: Statistics for Engineers and Scientists”, World Scientific Publishing Co. Pte. Ltd., 2006.
14. Reasenberg, R. D., Technical Memo TM10-05 (2010).
15. In a rather similar manner, the X-Y position of the scanner head for the formerly Park Scientific Instruments M5 and LS line of AFMs was servo-controlled independently from the main feedback loop of the Z-position.
16. Private communication with James P. Cowin, EMSL, Pacific Northwest National Laboratory.